\documentstyle[twocolumn,eqsecnum,aps,amssymb]{revtex}

\def\a{\alpha}
\def\b{\beta}
\def\g{\gamma}
\def\e{\varepsilon}
\def\th{\vartheta}
\def\s{\sigma}
\def\t{\tau}
\def\ph{\varphi}
\def\oo{\omega}
\def\O{\Omega}

\def\square{{\vcenter{\vbox{\hrule height.4pt \hbox{\vrule width.4pt 
height1.45ex \kern1.45ex \vrule width.4pt}
\hrule height.4pt}}}}

\def\qed{\penalty10000\hfill$\square$\par\goodbreak\medskip}

\def\proof{\medskip\noindent{\it Proof.} }

\def\NN{{\Bbb N}}
\def\RR{{\Bbb R}}
\def\PP{{\Bbb P}}
\def\EE{{\Bbb E}}

\def\tr{{\rm tr}\,}

\def\li#1{\lim_{#1\to\infty}}
\def\soi#1{\sum_{#1=1}^\infty}
\def\szi#1{\sum_{#1=0}^\infty}
\def\tuple#1_#2{#1_1,#1_2,\ldots,#1_{#2}}
\def\tup#1_#2{#1_1,\ldots,#1_{#2}}

\def\norm#1{\|#1\|}

\def\Er{\EE_\rho}
\def\Pr{\PP_\rho}
\def\Prt{\widetilde{\PP}_\rho}
\def\Ert{\widetilde{\EE}_\rho}
\def\Jnk{{\cal J}_{n,k}}

\begin{document}
\draft

\title{An Ergodic Theorem for Quantum Counting Processes}

\author{B.~K\"ummerer$^1$ and H.~Maassen$^2$}
\address{$^1$Mathematisches Institut A, 
         Universit\"at Stuttgart, 
         Pfaffenwaldring 57, 
         D-70569 Stuttgart, 
         Germany, 
         {\tt kuem@mathematik.uni-stuttgart.de}\\
         $^2$Mathematisch Instituut, 
         Katholieke Universiteit Nijmegen, 
         Toernooiveld 1, 
         6525 ED Nijmegen, 
         The Netherlands, 
         {\tt maassen@sci.kun.nl}}

\date{February 9, 2001}
\maketitle
\begin{abstract}
For a quantum-mechanical counting process we show ergodicity,
under the condition that the underlying open quantum system
approaches equilibrium in the time mean.
This implies equality of time average and ensemble average for
correlation functions of the detection current to all orders
and with probability 1.
\end{abstract}

\section{Introduction}\noindent
Modern research on quantum-mechanical counting processes,
be it numerical simulations
\cite{Carmichael} or experimental investigations
\cite{MabuchiYK},
usually starts from the tacit assumption that for the study of statistical
properties of the counting records it does not make a difference whether
a large number of experiments is performed or a single very long one. 
This assumption amounts to ergodicity of these records.
In several recent discussions, e.g. 
\cite{BriegelESW,NaraschewskiS,PlenioK,Cresser,DalibardCM},
investigators have addressed the question of its validity.
A partial result was obtained by Cresser \cite{Cresser},
who proved ergodicity in the $L^2$-sense and to first order 
in the detection current.
In this paper we establish ergodicity in the full sense (Theorem 3),
in particular to all orders in the detection current and with probability 1
(Theorem 4).
Theorem 5 formulates ergodicity in terms of multi-time coincidences.

For the description of detection records we employ the rigorous formulation 
of Davies and Srinivas \cite{Davies,SrinivasD},
which has set the tone for later investigations 
\cite{Carmichael,WisemanM,GardinerZ}.

\section{Counting processes according to Davies and Srinivas}\noindent
We consider an open quantum system under continuous observation by use of
a finite number $k$ of detectors.
The state of the system is described by a density matrix $\rho$ on
a Hilbert space, obeying a Master equation
$\dot\rho=L\rho$, where $L$ is a generator of Lindblad form \cite{Lindblad}.
Normalisation is expressed by the relation 

\begin{equation}
\tr L(\rho)=0\quad\hbox{for all }\rho\;.
\label{normal}
\end{equation}
A counting process connected to this quantum evolution is based on an
unraveling of the generator

\begin{equation}
   L=L_0+\sum_{i=1}^k J_i\;,
\label{unravel}
\end{equation}
which is interpreted as follows.
The reaction of the detectors to the system consists
of clicks at random times.
The evolution $\rho\mapsto e^{tL_0}(\rho)$ denotes the change of the state of the system
under the condition that during a time interval of length $t$
no clicks are recorded.
The operator $\rho\mapsto J_i(\rho)$ on the state space
describes the change of state conditioned on the occurrence of a click of
detector $i$.
For computational convenience we assume these operators to be bounded.
So, if $\rho$ describes the state of the system at time 0,
and if, during the time interval $[0,t]$, clicks are recorded
at times $\tuple t_n$ of detectors $\tuple i_n$ respectively,
and none more,
then, up to normalisation, the state at time $t$ is given by 

\begin{equation}
e^{(t-t_n)L_0} J_{i_n} e^{(t_n-t_{n-1})L_0}\cdots e^{(t_2-t_1)L_0}
                       J_{i_1} e^{t_1L_0}(\rho)\;.
\label{probab}
\end{equation}
The probability density $f^t((t_1,i_1),\ldots,(t_n,i_n))$
for these clicks to occur is equal to the trace of (\ref{probab}).

We imagine the experiment to continue indefinitely. The observation
process will then produce an infinite detection record
$\bigl((t_1,i_1),\ (t_2,i_2),\ (t_3,i_3),\ \dots\bigr)$\ ,
where we assume that
$0 \le t_1 \le t_2 \le t_3 \le \dots$\ , and $\lim_{n\to\infty} t_n = \infty$
(i.e., the clicks do not accumulate).

Let $\Omega$ denote the space of all such detection records.
By an {\it event} we mean some property of the record,
which we identify with the set $E\subset\O$ 
of all records with this property.
The events decidable at or before time $t\ge0$
form a $\sigma$-algebra $\Sigma_t$ \cite{Davies}.
Together these $\sigma$-algebras generate
the full $\sigma$-algebra $\Sigma$.
Following Davies and Srinivas we may now formulate the effect of
observation on the quantum system as follows:
If $t$ is a positive time,
$E$ an event in $\Sigma_t$, and
$\rho$ denotes a state, then we define

\begin{eqnarray}
&&M_t(E)(\rho):=\nonumber\\
&&\sum_{n=0}^\infty \sum_{i_1=1}^k \dots\sum_{i_n=1}^k
                 \int_0^t\int_0^{t_n}\cdots\int_0^{t_2}
                 1_E\bigl((t_1,i_1),\ldots,(t_n,i_n)\bigr)\nonumber\\
    &&e^{(t-t_n)L_0}J_{i_n} e^{(t_n - t_{n-1})L_0} \dots e^{(t_2 - t_1)L_0}
                J_{i_1}e^{t_1L_0}(\rho)\nonumber\\
&&\qquad\qquad\times dt_1dt_2\cdots dt_n\;.
\label{defMt}
\end{eqnarray}
Here $1_E$ denotes the indicator function of the event $E$ and
$M_t(E)$ is the effect on the quantum system of the occurrence of 
$E \in \Sigma_t$. Then
          
\begin{equation}
\PP_\rho^t(E) := \tr M_t(E)(\rho)
\label{defP}
\end{equation}
is the probability of the occurrence of
$E$ given that the system starts in $\rho$.
We extend the notation (\ref{defP}) also to density matrices $\rho$
which are not normalised.
The counting process as a whole is described by the family $(M_t)_{t\ge0}$.
The effect of the counting on the quantum system, 
when the outcome is ignored, is the time evolution
\begin{displaymath}
T_t(\rho):=M_t(\Omega)(\rho)\;.
\end{displaymath}
It follows from the Dyson series (\ref{defMt}) with $E=\O$
that $T_t$ is indeed the original time evolution
$e^{tL}$, in particular, by (\ref{normal}), $T_t$ preserves the trace.

\section{Ergodic theory}\noindent
The time shift by $t$ seconds is described by the map $\sigma_t$ on $\Omega$,
which is given on a particular record 
$\omega =((t_1,i_1),\ (t_2,i_2),\ (t_3,i_3),\ \dots) \in \Omega$
with $t_k \le t < t_{k+1}$ by
$\sigma_t(\omega) := ((t_{k+1}-t,i_{k+1}),  (t_{k+2}-t,i_{k+2}), \dots  )$.
The time shift of an event $E$ towards the future is given by $\s^{-1}_t(E)$.

The crucial property of the counting process $(M_t)_{t\ge0}$ is the following.
For all $s,t \ge 0$ and all events $E \in \Sigma_s$, $F \in \Sigma_t$ we have
\begin{equation}
M_{s+t}(F\cap\s^{-1}_t(E)) = M_s(E) \circ M_t(F) \ .
\label{Markov}
\end{equation}
This Markov property was proved in \cite{Davies}.
Putting $E=F=\O$ we recover the semigroup property $T_{s+t}=T_s\circ T_t$
of the time evolution.

When $F \in \Sigma_t$ and $s \ge 0$ then $\PP^{t+s}_\rho(F)$ does not
depend on $s$.
Indeed, since $\O=\s_t^{-1}(\O)$ and $T_s$ preserves the trace,

\begin{eqnarray}
\PP_\rho^{t+s}(F)
               &=& \tr\bigl(M_{t+s}(F)(\rho)\bigr) 
                = \tr\bigl(M_{t+s}(F\cap \s_t^{-1}(\Omega))(\rho)\bigr)
                  \nonumber\\
&\stackrel{\mathrm{(\ref{Markov})}}{=}&
                  \tr\bigl(M_s(\Omega)\circ M_t(F)(\rho)\bigr)
                = \tr\bigl(T_s\circ M_t(F)(\rho)\bigr)\nonumber \\
               &=&\tr\bigl(M_t(F)(\rho)\bigr) = \PP_\rho^t(F)\;.\nonumber
\end{eqnarray}
Therefore, by Kolmogorov's extension theorem,
the family $(\PP_\rho^t)_{t\ge0}$ of probability measures on the
$\sigma$-algebras $(\Sigma_t)_{t\ge0}$ with densities
$(f^t)_{t\ge0}$ extends to a single probability measure $\PP_\rho$
on the full $\s$-algebra $\Sigma$.

\proclaim Lemma 1.
For all $t\ge0$, all $E\in\Sigma$, $F\in\Sigma_t$, and all states $\rho$:
  \begin{equation}
\PP_\rho(F\cap \s_t^{-1}(E))=\PP_{M_t(F)(\rho)}(E)\;.
\label{PPF}
\end{equation}
In particular,
  \begin{equation}
\label{PPE}
\PP_\rho(\s_t^{-1}(E))=\PP_{T_t\rho}(E)\;.\end{equation}
Therefore, if $\rho$ is invariant under $T_t$, then $\PP_\rho$ is a
stationary probability measure on $\O$.

\proof
First suppose that $E\in\Sigma_s$.
Equality (\ref{PPF}) is obtained from the Markov property (\ref{Markov})
by acting on $\rho$ and taking the trace on both sides.
(\ref{PPE}) follows by putting $F=\O$.
The statements extend to all $E\in\Sigma$
by Kolmogorov's extension theorem since $s$ was arbitrary.
\qed
\medskip\noindent{\bf Definition.}
\begin{itemize}
\item
The evolution $(T_t)_{t\ge0}$ of a quantum system is said to
{\it converge in the mean} to an {\it equilibrium state} $\rho$
if for all normalised density matrices $\th$ and all observables $x$:
   \begin{displaymath}
\li \t{1\over\t}\int_0^\t \tr\bigl((T_t\th)x\bigr) \,dt=\tr(\rho x)\;.\end{displaymath}
\item
The counting process $(M_t)_{t\ge0}$ will be called {\it ergodic}
if the following holds.
Given any time-invariant event $E$, i.e. $\s^{-1}_t(E)=E$ for all $t\ge0$,
then either $\PP_\th(E)=0$ for all density matrices $\th$
or $\PP_\th(E)=1$ for all $\th$.
\end{itemize}
The condition on $(T_t)_{t\ge0}$ is satisfied in many cases of practical
importance.

\proclaim Theorem 2.
If the evolution $T_t=e^{tL}$, $t\ge0$, converges in the mean,
then the counting process $(M_t)_{t\ge0}$ is ergodic for any unraveling
(\ref{unravel}).

\proof
Let $E$ be a time-invariant event and $\th$ any state.
Then by (\ref{PPE}),
$\PP_\th(E)=\PP_\th(\s^{-1}_t(E))=\PP_{T_t\th}(E)$.
Since $\PP_\th$ is linear and continuous in $\th$,
we may average both sides
over the interval $[0,\t]$ and take the limit $\t\to\infty$ to obtain
$\PP_\th(E)=\PP_\rho(E)$.
For an unnormalised density matrix $\chi$ we find instead that
\begin{equation}
\PP_\chi(E)=\PP_\rho(E)\tr(\chi).
\label{rhoeqth}
\end{equation}
If $F$ is any event in $\Sigma_t$ then

\begin{eqnarray}
\PP_\th(F\cap E)&=&\PP_\th(F\cap \s^{-1}_t(E))
\stackrel{\mathrm{(\ref{PPF})}}{=}\PP_{M_t(F)(\th)}(E)\nonumber\\ 
&\stackrel{\mathrm{(\ref{rhoeqth})}}{=}&
                     \PP_{\rho}(E)\,\tr(M_t(F)(\th))
                    =\PP_{\rho}(E)\PP_\th(F)\nonumber \\
&\stackrel{\mathrm{(\ref{rhoeqth})}}{=}&
                     \PP_{\th}(E)\PP_\th(F)\;.\nonumber
\end{eqnarray}
The resulting equation extends to all $F\in\Sigma$,
in particular it holds for $F=E$:
   \begin{displaymath}
\PP_\th(E)=\PP_\th(E)^2\;.\end{displaymath}
It follows that $\PP_\th(E)$ is equal to 0 or 1.
\qed

Let us denote the expectation $\int_\O f(\oo)d\Pr(\oo)$
of an integrable function $f$ on $\O$ by $\Er(f)$.

\proclaim Theorem 3.
If the evolution $(T_t)_{t\ge0}$ converges in the mean to $\rho$,
then for all integrable functions $h$ on $\Omega$
and all initial states $\th$ we have, almost surely with respect to $\PP_\th$,
   \begin{equation}
\li\t{1\over\t}\int_0^\t h(\s_t(\oo))\,dt=\EE_\rho(h)\;.
\label{erg}
\end{equation}

\proof
By Lemma 1 and Theorem 2, $\PP_\rho$ is stationary and ergodic.
Hence, by Birkhoff's individual ergodic theorem, the limit on the left exists
almost surely with respect to $\PP_\rho$,
and is equal to the constant $\EE_\rho(h)$.
Since the set $F$ of points $\oo\in\O$ for which (\ref{erg}) holds,
is time-invariant, we have $\PP_\th(F)=\PP_\rho(F)=1$
for all states $\th$ by (\ref{rhoeqth}).
\qed

\section{Applications}\noindent
The main result of the present ergodic theory for quantum counting processes,
Theorem 3,
can be made considerably more concrete by applying it to detection
currents and multi-time coincidences, showing bunching or anti-bunching.

For simplicity we consider only one detector,
which responds to a point event at time $s$ by producing a current
$\g(t-s)$ at time $t$.
(This will be zero for $t<s$.)
The total detection current is given by
\begin{displaymath}
I_t(\oo):=\sum_{s\in\oo}\g(t-s)\;.
\end{displaymath}

Let $\Prt$ be the unique stationary extension of $\Pr$ to
negative times on
the configuration space $\widetilde\O$ of the full real line.
We shall denote expectation with respect to this measure by $\Ert$.

\proclaim Theorem 4.
Let the quantum evolution $(T_t)_{t\ge0}$ converge in the mean
to a state $\rho$
and let the detector response function $\g:\RR\to[0,\infty)$
be bounded and integrable.
Then for all $0\le t_1\le t_2\le\ldots\le t_n$ and all initial
states $\th$ we have, almost surely with respect to $\PP_\th$,
\begin{displaymath}
\li\t{1\over\t}\int_0^\t I_{t_1+t}(\oo)\cdots I_{t_n+t}(\oo)\,dt
            =\Ert\left(I_{t_1}\cdots I_{t_n}\right)\;.
\end{displaymath}

For $n=2$ this theorem implies a quantum-mechanical version of the
Wiener-Khinchin theorem.
In the proof we shall make use of the non-exclusive probability density
of the stationary process \cite{vanKampen,GardinerZ,Cresser},

\begin{displaymath}
g_n(\tuple t_n)
    :=\tr\left(JT_{t_n-t_{n-1}}J\cdots JT_{t_2-t_1}J(\rho)\right)\;.
\end{displaymath}

The functions $g_n$ are related to the probability density $f^t$ from
(\ref{probab}) of the counting process (where $t\ge t_n$), by
\begin{eqnarray}
&&g_n(\tuple t_n)
  =f_n^t(\tuple t_n)\nonumber \\
  &+&\soi m\int_0^t\int_0^{s_m}\cdots\int_0^{s_2}
  f^t_{m+n}\bigl(\{\tup t_m\}\cup\{\tup s_n\}\bigr)\nonumber\\
        &&\times ds_1\cdots ds_m
  =\int_{\O_t} f^t(\{\tuple t_n\}\cup\oo)d\oo\;;
\label{gn}
\end{eqnarray}
here $\O_t$ is the set of finite subsets of $[0,t]$,
which can be identified with the time-ordered points in
$\{\emptyset\}\cup\bigcup_{m=1}^\infty[0,t]^m$.
By $d\oo$ we mean $ds_1ds_2\cdots ds_m$ if $\oo=\{\tuple s_m\}$ with
$s_1\le s_2\le\cdots\le s_m$.

\medskip\noindent{\it Proof of Theorem 4}.
First we note that Theorem 3 also holds if $\O$, $\Pr$ and $\Er$ are
replaced by
$\widetilde\O$, $\Prt$ and $\Ert$ respectively,
as introduced above,
and $\s_t$ by the left shift of $\oo\subset\RR$.
Then we have
$I_{s+t}(\oo)=I_s(\s_t(\oo))$.
Now fix $n\in\NN$ and $0\le t_1\le\cdots\le t_n$.
Let $h:\widetilde\O\to\RR$ be given by
\begin{displaymath}
h(\oo):=I_{t_1}(\oo)I_{t_2}(\oo)\cdots I_{t_n}(\oo)\;.
\end{displaymath}
It follows that
$h\circ\s_t=I_{t_1+t}I_{t_2+t}\cdots I_{t_n+t}$,
and the statement to be proved follows from Theorem 3,
provided that $h$ is integrable.
In the Appendix we shall show that this is indeed the case.
\qed

\smallskip
As our second application we shall show
that the non-exclusive probability densities $g_n$ have a
straightforward pathwise interpretation:
they are equal to the frequency of multi-time coincidences on almost every
detection record.
For this, let $N_{[a,b]}(\oo):=\#(\oo\cap[a,b])$ denote the number of clicks detected 
during the time interval $[a,b]$.

\proclaim Theorem 5.
Let $(T_t)_{t\ge0}$ converge in the mean to the equilibrium state $\rho$.
Then for all $n\in\NN$, all $0\le t_1\le t_2\le\ldots\le t_n$,
all $\e$ between $0$ and $\min_{1\le j<n}(t_{j+1}-t_j)$, and all initial
states $\th$ we have, almost surely with respect to $\PP_\th$,
\begin{eqnarray}
&&\li\t{1\over\t}\int_0^\t
     \left(\prod_{j=1}^n N_{[t_j+t,t_j+t+\e]}(\oo)\right)dt\nonumber\\
   &&=\int_{t_n}^{t_n+\e}\cdots\int_{t_1}^{t_1+\e} g(\tup s_n)
         ds_1\cdots ds_n\;.
\end{eqnarray}

\proof
Fix $n\in\NN$ and a sequence
$0\le t_1\le t_2\le\cdots\le t_n$ of times.
Let $K:\O\to\{0,1\}$ be the function that maps $\oo\in\O$ to 1 if $\oo$
contains
exactly $n$ points, one in each of the intervals
$[t_1,t_1+\e],\cdots,[t_n,t_n+\e]$, and to $0$ otherwise.
Then we obtain for $t\ge t_n+\e$, using set notation and the integral-sum
lemma from \cite{LindsayM},

\begin{eqnarray}
           &\displaystyle\int_{t_n}^{t_n+\e}&\cdots\int_{t_1}^{t_1+\e}
                    g(\tup s_n)\,ds_1\cdots ds_n\nonumber\\
            &=&\int_{\O_t} K(\a)g(\a)d\a
\stackrel{\mathrm{(\ref{gn})}}{=}\int_{\O_t}\int_{\O_t}
                      K(\a)f^t(\a\cup\b)d\a d\b\nonumber\\
&\stackrel{\mathrm{\cite{LindsayM}}}{=}&
               \int_{\O_t}\left(\sum_{\a\subset\oo}K(\a)\right)f^t(\oo)d\oo.
\label{intsumK}
\end{eqnarray}
A short calculation shows that
\begin{equation}
\sum_{\a\subset\oo}K(\a)=\prod_{j=1}^n N_{[t_j,t_j+\e]}(\oo).
\label{sumK}
\end{equation}
Since $0\le g_n(\tuple s_n)\le\norm{J}^n$, the integral (\ref{intsumK})
is convergent,
hence the product on the r.h.s. of (\ref{sumK}) is integrable as a function
of $\oo$. 
Application of Theorem 3 to this product now yields the statement.
\qed

\section{Discrete time}\noindent
There is an obvious analogue of our main result (Theorem 3)
in discrete time \cite{MaassenK}.
A Kraus measurement \cite{Kraus} is given by a decomposition of a 
completely positive operator $T$ on state space as
\begin{displaymath}
T\rho=\sum_{i=1}^k a_i\rho a_i^*\;,
\end{displaymath}
where $\rho\mapsto a_i\rho a_i^*$ describes the state change of the
density matrix $\rho$ when the measurement gives the outcome $i$.
Thus for initial state $\th$ the probability of finding the sequence
of outcomes $\tuple i_m$ by repeated Kraus measurement is given by
\begin{displaymath}
\tr\left(a_{i_m}\cdots a_{i_1}\th a_{i_1}^*\cdots a_{i_m}^*\right)\;.
\end{displaymath}
As in continuous time, this yields a probability measure $\PP_\th$
on the space of detection records $\O:=\{1,2,\cdots,k\}^\NN$.
Again, if $(T^n)_{n\in\NN}$ converges in the mean to some state $\rho$,
then the only time invariant events in $\O$ have measure 0 or 1 for all $\PP_\th$.
In particular, $\Pr$ is ergodic.

\appendix
\section*{}
\noindent
We shall show that, in the situation of Theorem 4, $h:=I_{t_1}\cdots I_{t_n}$
is an integrable function on $\widetilde\O$ provided that the jump operator
$J$ is bounded and the detector response function $\g:\RR\to[0,\infty)$ is
bounded and integrable. 

Let $M:=\max(1,\norm\g_\infty)$.
Fix $n\in\NN$ and a sequence $0\le t_1\le t_2\le\ldots\le t_n$ of times.
Let
\begin{displaymath}
\ph(t):=\sum_{j=1}^n\g(t_j-t)\;.
\end{displaymath}
Then $\ph$ is also integrable, with $\norm\ph_1=n\norm\g_1$.
For $k\in\NN$, let $\Jnk$ denote the set of all surjections 
$\{1,\cdots,n\}\to\{1,\cdots,k\}$.
Then we may write for any $\oo\in\widetilde\O$,

\begin{eqnarray}
&I_{t_1}(\oo)&I_{t_2}(\oo)\cdots I_{t_n}(\oo)\nonumber\\
      &=&\sum_{s_1\in\oo}\cdots\sum_{s_n\in\oo}\g(t_1-s_1)\cdots\g(t_n-s_n)\nonumber \\
      &=&\sum_{k=1}^n\sum_{j\in\Jnk}
         \sum_{{\{\tup a_k\}\subset\oo}\atop{a_1<\ldots<a_k}}
            \g(t_1-a_{j(1)})\cdots\g(t_n-a_{j(n)})\nonumber \\
      &\le&\sum_{k=1}^n\#\left(\Jnk\right)\sum_{{\a\subset\oo}\atop{\#\a=k}}
            {\norm\g}_\infty^{n-k}\left(\prod_{s\in\a}\ph(s)\right)\nonumber \\
&\le& n\cdot n^n M^n\sum_{\a\subset\oo}\left(\prod_{s\in\a}\ph(s)\right).
\label{prodI}
\end{eqnarray}
Using set notation and the integral-sum lemma \cite{LindsayM} again
we conclude that, for all $t\ge0$ and $u\ge t_n+t$,

\begin{eqnarray}
\Er&&\bigl((I_{t_1}I_{t_2}\cdots I_{t_n})\circ\s_t\bigr)/M^nn^{n+1}\nonumber\\
&&\stackrel{\mathrm{(\ref{prodI})}}{\le}\int_{\O_u}\sum_{\a\subset\oo}
             \left(\prod_{s\in\a}\ph(s-t)\right)f^u(\oo)\,d\oo\nonumber \\
&&\stackrel{\mathrm{\cite{LindsayM}}}{=}\int_{\O_u}\int_{\O_u}
          \left(\prod_{s\in\a}\ph(s-t)\right)f^u(\a\cup\b)\,d\a d\b\nonumber \\
&&\stackrel{\mathrm{(\ref{gn})}}{=}
          \int_{\O_u}\left(\prod_{s\in\a}\ph(s-t)\right)g(\a)d\a\nonumber \\
&&\le\szi m{{\norm J^m}\over{m!}}\int_{[0,u]^m}\ph(s_1-t)\cdots\ph(s_m-t)
          ds_1\cdots ds_m\nonumber \\
        &&\le\exp\left(\norm J\int_0^u\ph(s-t)ds\right)
         \le e^{n\norm J\cdot\norm\g_1}.   \nonumber
\end{eqnarray}
Therefore, since the r.h.s. does not depend on $t$,
\begin{displaymath}
\Ert(I_{t_1}\cdots I_{t_n})
  =\li t\Er\bigl((I_{t_1}\cdots I_{t_n})\circ\s_t\bigr)<\infty.
\end{displaymath}
\qed

\acknowledgments\noindent
Support from the `Deutsche Forschungsgemeinschaft' (DFG) and
`Stichting Fundamenteel Onderzoek der Materie' (FOM) is 
gratefully acknowledged.


\end{document}